\documentclass[11pt, a4paper]{article}

\usepackage[pdftex]{graphicx}
\usepackage{bm,afterpage,amsmath,placeins,flafter,authblk,adjustbox,lscape,caption}
\usepackage[authoryear]{natbib}
\usepackage[english]{babel}
\usepackage{setspace}
\usepackage{booktabs}
\usepackage[top=1.5in,left=1.25in,right=1.25in,bottom=1.5in]{geometry}
\usepackage{color}
\usepackage{hyperref}
\graphicspath{{figures/}}

\date{}

\begin{document}

\title{\textbf{Spatio-Temporal Movements in Team Sports: A Visualization approach using Motion Charts.}}

\newcommand{\authorlastnames}{Metulini - Sports' Movements using Motion Charts}

\author[a]{Rodolfo Metulini
	\thanks{Corresponding authors: %
		rodolfo.metulini@unibs.it.}}


\affil[a]{ Department of Economics and Management, University of Brescia \\ Contrada Santa Chiara, 50, 25122 Brescia BS, Italy.}

\date{ [Submitted to a Journal. For citation, please contact the author] }



 \maketitle 

\begin{abstract}

To analyze the movements and to study the trajectories of players is a crucial need for a team when it looks to improve its chances of winning a match or to understand its performances. State of the art tracking systems now produce spatio-temporal traces of player trajectories with high definition and frequency that has facilitated a variety of research efforts to extract insight from the trajectories. Despite many methods borrowed from different disciplines (machine learning, network and complex systems, GIS, computer vision, statistics) has been proposed to answer to the needs of teams, a friendly and easy-to-use approach to visualize spatio-temporal movements is still missing. This paper suggests the use of \texttt{gvisMotionChart} function in \texttt{googleVis} \texttt{R} package. I present and discuss results of a basketball case study. Data refers to a match played by an italian team militant in "C-gold" league on March 22nd, 2016. With this case study I show that such a visualization approach could be useful in supporting researcher on preliminar stages of their analysis on sports' movements, and to facilitate the interpretation of their results.  
\paragraph{keywords:} Spatio-Temporal Movements, Trajectories, Sports Statistics, Motion Charts, GoogleVis.
\end{abstract}

\section{Introduction}\label{sec:intro}

Teams looking to improve their chances of winning will naturally seek to understand their performance, and also that of their opposition. From a data mining perspective, \citet{carpita2013football} and \citet{carpita2015discovering} used cluster and principal component analysis techniques in order to identify the drivers that most affect the probability to win a match.

From a different perspective, nowadays, analyzing players' trajectories using spatio-temporal data is becoming cool \footnote{Apart from sports, other fields of study have encountered the need to analyze trajectories in a space-time dimension. This is, for example, the case of animal movement \citep{brillinger2004exploratory,calenge2007exploring,calenge2009concept,schwager2007robust}}.  There are a number of IT systems in use that capture these data from team sports during matches. Spatio-temporal data are characterized by a sequence of samples containing the timestamp and location of some phenomena. In the team sports domain, two types of spatio-temporal data are commonly captured: object trajectories capture the movement of players or the ball; event logs record the location and time of match events, such as passes, shots at goal or fouls. The movement of players or the ball around the playing area are sampled as a timestamped sequence of location points in the plane. The trajectories are captured using optical- or device-tracking and processing systems. Optical tracking systems use fixed cameras to capture the player movement, and the images are then processed to compute the trajectories \citep{bradley2007reliability}. In other cases, device tracking systems rely on devices that infer their location, and are attached to the players' clothing or embedded in the ball. It is a common procedure to discretize the playing area into regions and assign the location points contained in the trajectory to a discretized region. A common approach is to subdivide the playing area into rectangles of equal size \citep{cervone2016multiresolution}.

Understanding the interaction between players is one of the more important and complex problem in sports science. Trajectories allow to analyze movements of a single player as well as interactions of all players as a synchronized group, in order to asses the importance of such players to the team. Methods used to analyze these movements using trajectories borrow from many disciplines, such as machine learning, network and complex systems, GIS, computational geometry, computer vision and statistics.

For example, the central goal of social network analysis is to capture the interactions between individuals \citep{wasserman1994social}. As a consequence, in the last decade numerous papers applied social network analysis to team sports, mainly focusing on passing networks and transition networks. Centrality techniques has been used with the aim of identifying key (or \textit{central}) players, or to estimate the interaction and the cooperation between team members \citep{passos2011networks}.

Furthermore, control space is considered a key factor in the team's performance. A player dominates an area if he can reach every point in that area before anyone else. Literature makes use of the tool of the dominant region \citep{taki1996development}, which is equivalent to the \textit{Voronoi} region \citep{fortune1987sweepline}, when acceleration is constant. Another approach regards measuring the average distance of the players in the court, and its evolution over time. Many works are devoted to analyze as the space is occupied by players - when attacking and when defending - or in crucial moments of the match. We can find examples in football \citep{couceiro2014dynamical,moura2012quantitative} or in futsal \citep{fonseca2012spatial,travassos2012spatiotemporal}

Another issue regards to model the evolution of football play from the trajectories of the players, which has been researched extensively, particularly in the computer vision community \citep{yue2014learning,wei2014forecasting}. For example \citet{kim2010motion} predicted the location of the ball at a point in the near future. 

Predefined plays are used in many team sports to achieve some specific objective: teammates who are familiar with each other's playing style may develop ad-hoc productive interactions that are used repeatedly. \citet{brillinger2007potential} addressed the question how to describe analytically the spato-temporal movement of particular sequences of passes (i.e. the last 25 passes before a score). Moreover, segmenting a match into phases is a common task in sports analysis, as it facilitates the retrieval of important phases for further analysis.  \citet{perin2013soccerstories} developed a system for visual exploration of phases in football.

\vline

As described above, a variety of approaches and methods has already been proposed to solve different issues related to the relation between trajectories and performances in team sports. This paper provides a simple and ad-hoc strategy to visualize the spatio-temporal movement of a player as well as the synchronized movements of all players of the team. The aim is to support researchers on preliminar stages of their analysis as well as to facilitate the interpretation of results. To this scope, I propose to use motion charts. A motion chart is a dynamic bubble chart which allows efficient and interactive exploration and visualization of multivariate data. Motion charts map variables into time, 2D coordinate axes, size and colors, and facilitate the interactive display of multidimensional and temporal data. The best known motion chart, popularised by Hans Rosling in his TED talks, is probably the one provided within \texttt{googleVis} package in R, \texttt{gvisMotionChart}. This function allows the user to visualize data stored in R data frames directly in a web browser.  

\vline

In section \ref{sec:motionchart} I introduce the \texttt{gvisMotionChart} function in \texttt{googleVis} package, and I discuss this method in relation to team sports. Section \ref{sec:case} presents a case study based on trajectories data of basketball players. In this section I empirically show how the use of \texttt{gvisMotionChart} could gives us clues for further analysis. Section \ref{sec:concl} concludes and suggests future developments.

\section{Motion Charts for team sports' movements using googleVis}\label{sec:motionchart}

When talking of a motion chart we can think at a dynamic bubble chart. A bubble chart is a type of chart that displays three dimensions of data. Each entity with is a triplet of associated data, which is plotted expressing two of the three values through the XY-axes and the third through its size. Bubble charts can facilitate the understanding of social, economical, medical, and other scientific relationships. Bubble charts can be considered a variation of the scatter plot, in which the data points are replaced with bubbles. 
Motion Chart allows efficient and interactive exploration and visualization of space-time multivariate data and provides mechanisms for mapping ordinal, nominal and quantitative variables onto time, 2D coordinate axes, size and colors which facilitate the interactive display of multidimensional and temporal data. Motion charts provide a dynamic data visualization that facilitates the representation and understanding of large and multivariate data. Using the familiar 2D Bubble charts, motion Charts enable the display of large multivariate data with thousands of data points and allow for interactive visualization of the data using additional dimensions like time, size  and color, to show different characteristics of the data.

The central object of a motion chart is a bubble. Bubbles are characterized by size, position and appearance. Using variable mapping, motion charts allow control over the appearance of the bubble at different time points. This mechanism enhances the dynamic appearance of the data in the motion chart and facilitates the visual inspection of associations, patterns and trends in space-time data.

The \texttt{gvisMotionChart} is a function of \texttt{googleVis} package \citep{gesmann2013package}  which reads a \textit{data.frame} object and creates text output referring to the Google Visualisation API. It can be included into a web page, or as a stand-alone page. The actual chart is rendered by the web browser in Flash \footnote{It does not work in all the browsers, but require Google.}.
The function generates a motion chart, that is a dynamic chart which is traditionally designed to explore several indicators over time. Motion charts are intensively used and publicized by Ans Roslin trought TED. \texttt{gvisMotionChart} is used in a wide range of topics, such as students learning processes. \citet{santos2012goal} used different visualization methods available in \textit{googleVis}; \citet{hilpert2011dynamic} is an example of works where motion charts are adopted as a visual instrument in linguistic and semantic studies of the dynamic of linguistic change over the time. Motion Charts are applied different subfields of economics, for example in finance, to visualize sales data in an insurance context \citep{heinz2014practical} and for the study of inequality and income \citep{saka2015inequality}. In \citet{santori2014application} motion charts was applied to aggregated liver transplantation data. Visualization of Water Quality Sampling-Events in Florida are analyzed by means of motion charts in \citet{bolt2015visualizing}.

\vline

Analytically, the \texttt{gvisMotionChart} function reads as follow:

\vline 

 \texttt{gvisMotionChart(data, idvar = "id", timevar = "time", xvar = " ", yvar = " ", colorvar = " ", sizevar = " ",date.format = "Y/m/d", options = list(), chartid)}

\vline 

where 

\begin{itemize}
\item \texttt{data} is a data.frame object. The data has to have at least four columns with subject name (\texttt{idvar}), time (\texttt{timevar}) and two columns of numeric values. Further columns, numeric and character/factor are optional. The combination of \texttt{idvar} and \texttt{timevar} has to describe a unique row. The column names of the \texttt{idvar} and \texttt{timevar} have to be specified. Further columns, if not specified by the other arguments (\texttt{xvar, yvar, colorvar, sizevar}) will be assumed to be in the order of the arguments.
\item \texttt{idvar} is a column name of data with the subject to be analysed.
\item \texttt{timevar} is a column name of data which shows the time dimension. The information has to be either numeric, of class date or a character which follows the pattern 'YYYYWww' (e.g. '2010W04' for weekly data) or 'YYYYQq' (e.g. '2010Q1' for quarterly data).
\item \texttt{xvar}: column name of a numerical vector in data to be plotted on the x-axis.
\item 	\texttt{yvar}: column name of a numerical vector in data to be plotted on the y-axis.
\item \texttt{colorvar}: column name of data that identifies bubbles in the same series. Use the same value to identify all bubbles that belong to the same series; bubbles in the same series will be assigned the same color. Series can be configured using the \texttt{series} option.
\item \texttt{sizevar}: values in this column are mapped to actual pixel values using the sizeAxis. 
\item \texttt{options}: list of configuration options for Google motion chart. The options are documented in detail by Google online, 
\end{itemize}

\vline

Now, I contextualize the use of \texttt{gvisMotionChart} for the teams sports' movement.  Let suppose having data about a number of players: in our \textit{data.frame} object we should have a variable that uniquely identifies these players. This is the \texttt{idvar} variable. Our \textit{data.frame} should also contains a variable uniquely identifying the time dimension in which players' movements are tracked; this is the \textit{timevar} variable. A record in the \textit{data.frame} should be uniquely identified by the combination of \textit{idvar} and \textit{timevar}.
Moreover, our \textit{data.frame} should contains two additional variables containing the input for the x-axis and the y-axis. For the x-axis we have the position of the player in (let say) the court length and for the y-axis the position in (let say) the court width. These are, respectively, \texttt{xvar} and \texttt{yvar}.

\begin{figure}[!htb]
	\centering
	\includegraphics[width=0.65\textheight]{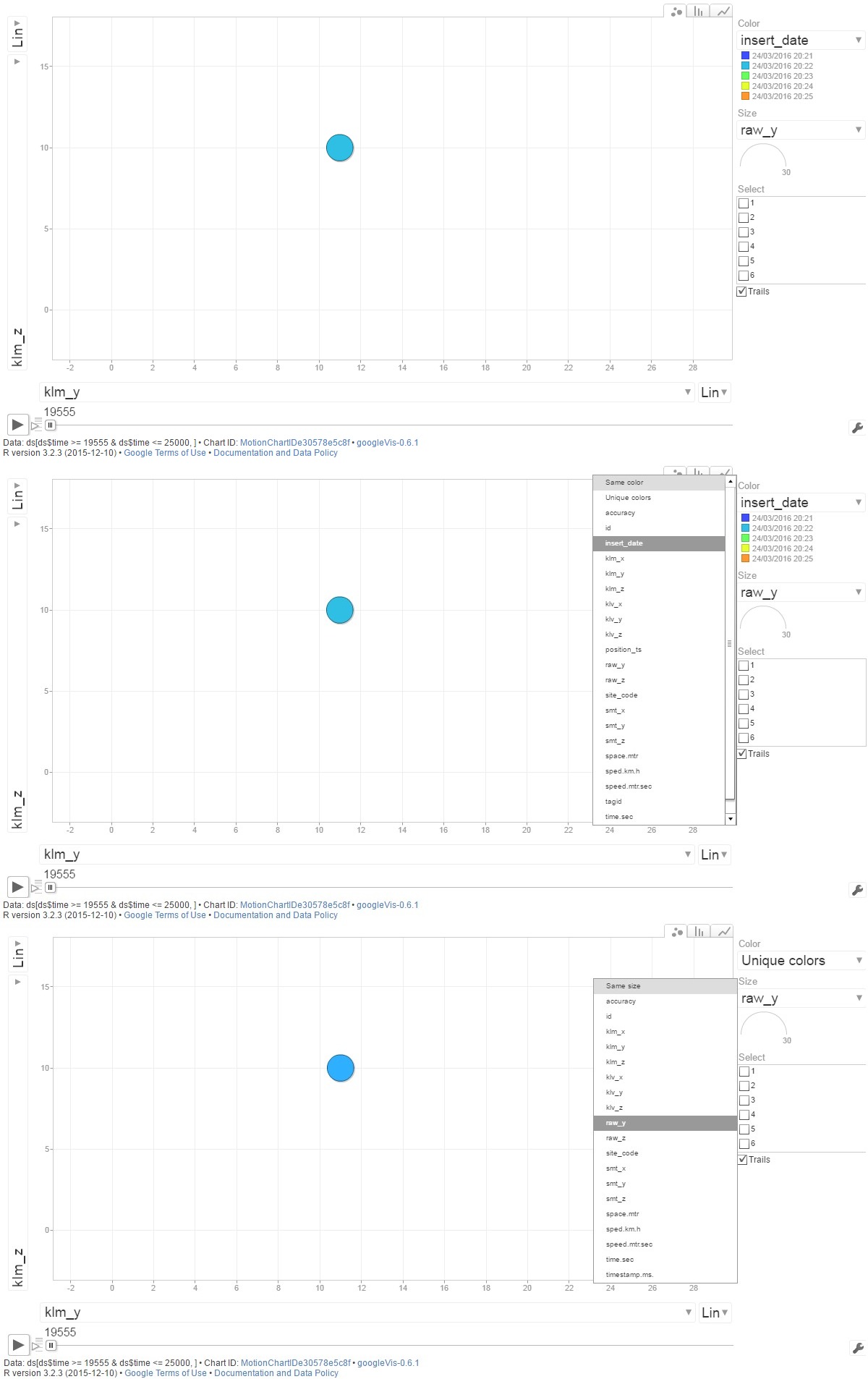}
	\caption{Setting the Motion Chart via html - 1}
	\label{mchart1}
\end{figure}

\begin{figure}[!htb]
	\centering
	\includegraphics[width=0.65\textheight]{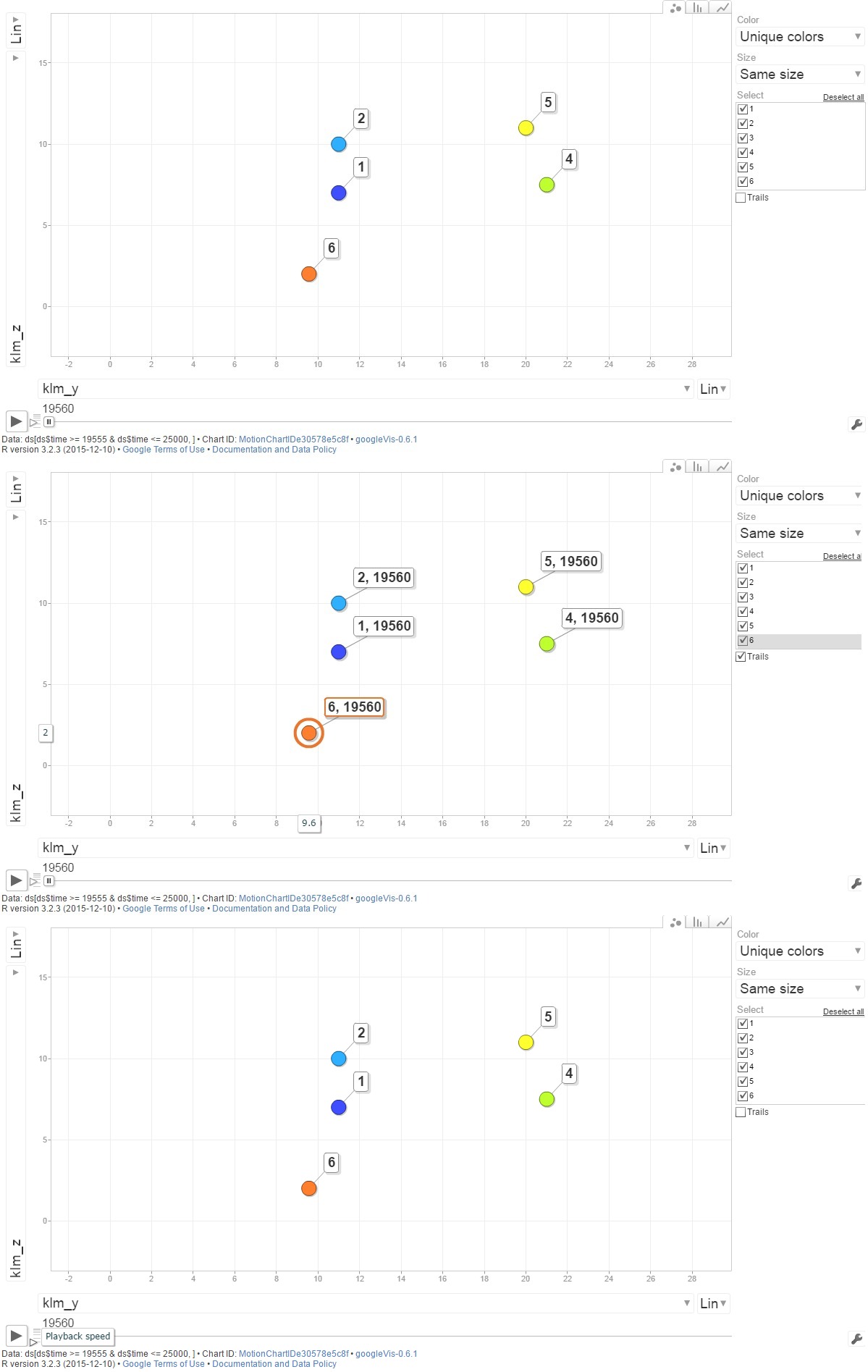}
	\caption{Setting the Motion Chart via html - 2}
	\label{mchart2}
\end{figure}

The \texttt{plot} function can be used as to represent the \texttt{googleVis} motion chart using browser. 

\texttt{Options} command can be used to define the court's dimension. By default, \texttt{gvisMotionChart} displays a squared chart (i.e. same length and width). With \texttt{Options} we can transform the court to be rectangular and with the right proportions. 

Summarizing, a \textit{data.frame} containing the four variables above described permits to visualize the dynamic of more than one player together: Different players can be reported with different (unique) colors (please see the middle chart in figure \ref{mchart1}).
Other variables should be supplied to the function in order to, for example, set the bubbles' dimension (see the bottom chart in figure \ref{mchart1}): when having available the x-axis and y-axis coordinates in successive moments of time, it is easy to compute the speed. A speed variable should be inputed to characterize the bubbles' dimension.  

It is possible to visualize the movement of one or more players together by ticking them in the appropriate box in the browser (top chart of figure \ref{mchart2}). In the same vein, we can activate players' trails: it will leave a line in the chart as bubbles play over time (middle chart of figure \ref{mchart2}). Finally, it is possible to set the speed by regulating the \textit{playback speed} key (please refer to bottom chart of figure \ref{mchart2}).

\section{Case Study} \label{sec:case}

Basketball is a sport generally played by two teams of five players each on a rectangular court. The objective is to shoot a ball through a hoop 18 inches (46 cm) in diameter and mounted at a height of 10 feet (3.05 m) to backboards at each end of the court. This sport was invented in Springfield (Canada) in 1891 by Dr. James Naismith. 

Rules of european basketball from FIBA (www.fiba.com) differs from the rules of the United States first league, National Basketball Association (NBA). The match lasts 40 minutes, divided in four periods of 10 minutes each. There is a 2 minutes break after the first quarter and after the third quarter of the match.  After the first half, there are 15 minutes of half-time break.  

This case study refers to a friendly match played on march 22th, 2016 by a team based in the city of Pavia (near Milano) called \textit{Winterass Omnia Basket Pavia}. This team, in the season 2015-2016, played in the "C gold" league, the fourth league in Italy. This league is organized in 8 divisions in which teams geographically close by play together. Each division is composed by 14 teams that play twice with every other team of the same division (one as a guest and one as a host team) for a total of 26 games in the regular season. At the end of the regular season, the top 8 teams in the final rank play a post season (also called "playoff") that serves as to declare the winning team as well as to determine the team that goes to the upper league in the next season. 

\subsection{Dataset description}

On march 22th, 2016, six \textit{Winterass} players took part of the friendly match. All that players have worn a microchip in their neck. The microchip tracks their movements in the court. The court length measures 28 meters while the court width equals to 15 meters.  The system collects the position (in pixels of 1 $m^2$) in both the two axis (respectively x-axis and y-axis), as well as in the z-axis (i.e. how much that player jumps). The positioning of the players has been detected for a huge amount of close instants of time measured in milliseconds. Considering all the six players, the system recorded a total of 133,662 space-time observations. More in detail, a list of collected data follows\footnote{A reduced version of the full dataset is available upon request.}:    

\begin{itemize}
	\item \textbf{id}: this is a ID variable that is unique for each record in the dataset.
	\item  Both \textbf{insert\_date} and \textbf{position\_ts} reports the date (dd/mm/aaa) and the time (hh:mm) of the detection.
	\item The column \textbf{tagid} uniquely identifies the player. In the dataset, 6 different ID are present, associated to each of the 6 players.
	\item \textbf{timestamp\_ms\_ok} reports the timestamp of the observation, in terms of milliseconds. 
	\item \textbf{smt\_x},\textbf{ smt\_y}, \textbf{smt\_z} reports the non filtered values for the x-axis, y-axis and z-axis.
	\item \textbf{klm\_x}, \textbf{klm\_y}, \textbf{klm\_z} instead, are the values for the x-axis, y-axis and z-axis, filtered with a Kalman approach.
	\item \textbf{klv\_x}, \textbf{klv\_y}, \textbf{klv\_z} reports the speed along, respectively, the x-axis, the y-axis and the z-axis, based on filtered data described above.
\item \textbf{tagid\_new} reports the same info of \textbf{tagid}, but here players are identified as 1, 2, ... , 6. 
\item \textbf{time} is a ID variable for the time dimension (i.e. the first record in terms of time is marked with a 1, the second record in terms of time with 2, etc...)  
\item \textbf{speed.mtr.sec}: It is a \textit{raw} measure of speed (in \textit{m/s}) of each player in each moment of time.
\end{itemize}

With regards to the same match, a play-by-play dataset is also available.  The play by play accounts for the actions of an event. 
In details, I have recorder the following variables: 

\begin{itemize}
\item \textbf{timestamp}: this variable reports the date (dd/mm/aaa) and the time (hh:mm) of the detection.
\item \textbf{action} is a string variable that reports the type of action (for example "two shot made", "rebound", ...).
\item \textbf{name}: The first name of the player to which the action is associated.
\item \textbf{surname}: The surname of the player to which the action is associated.
\item \textbf{x\_coord} and \textbf{y\_coord} report the x-axis and the y-axis (expressed in values from -100 to 100). The coordinate (0,0) is the center of the court.
\end{itemize}

Both the movements data and the play-by-play were kindly provided by MYagonism (\url{https://www.myagonism.com/}).
Unfortunately, the finest disaggregation level of the time (minutes) in the play-by-play does not permit to properly match the play-by-play with the movements data. So, in the following, the most part of the analysis only make use of the informations coming from the movements' dataset.

\subsection{Descriptive statistics}

Six player took turns in the court. The total number of records equals to 133,662.  Using the klm\_y and the klm\_z variables as the x- and y-axis and looking to the players' position in the court, I drop the pre-match, the half-time break and the post match periods from the full dataset (please refer to appendix \ref{AppA}). I end up with a total of 106,096 observations.  Having available the timestamp variable, I found that the match lasts for 3,971,180 milliseconds, which equals to about 66 minutes. This also mean that, in average, the system collects positions about 37 times every second (3,971,180 / 106,096). Considering that 6 players are in the court at the same time, the position of each single player is collected, in average, 6.2 times every second (in other words, the position of each player is collected, in average, every 162 milliseconds). 

\vline

Inspecting the data, I found that, out of the total of 106,096 observations, 17379 report the position of \textit{player 1}, 16708 report the position of \textit{player 2}, 15702 belong to \textit{player 3} while 18573 belong to \textit{player 4}. Moreover, \textit{player 5}'s and \textit{player 6}'s positions are collected, respectively, 18668 and 19066 times. This does not mean that the last three players remained in the court more than the others.  

Table \ref{summary} reports the summary statistics of the variables contained in the \textit{data.frame}.   	Min/max, mean and relevant quartiles are reported.  It emerged that players moves inside the area covering all the 1 m$^2$ cells in which the court has been divided; more in detail, players also stay the cells related to the bench area (when smt\_y and klm\_y reports negative values) as well as outside the court (it happen when the y-axis values are outside the interval [0,15] and when the x-axis values are outside the interval [0,28])
I note that filtered (with kalman) and not filtered coordinates are really close each other. The average x-axis value (length) equal to 12.67 and the average y-axis (width) to 6.29. Roughly the same when considering the filtered values. The z-axis (i.e. the height) values range from 0 to 3, meaning that the maximum height that a player reach is around three meters.
Speed (expressed in m/s) reports a mean of 1.89. 
 
\begin{table}[htbp]
	\centering
	\caption{Summary statistics for the relevant variables in the dataset}
	\small
	\begin{tabular}{l|rrrrrrrrrr}
&	smt\_z &	smt\_x &	smt\_y &	klm\_z	& klm\_x	& klm\_y	 & klv\_z &	klv\_x &	klv\_y	& speed(m/s) \\
\hline
Min. &	0.00 &	-2.00	&-2.00&	0.00	&-2.00&	-2.00&	0.00	&-5.00 &	-4.00&	0.00 \\
1st Qu. &	0.00 &	5.00&	4.00&	0.00	&5.00&	4.00&	0.00	&0.00&	0.00  & 0.00 \\
median & 0.00 &	11.00&	7.00&	0.00	&11.00&	7.00	&0.00	&0.00&	0.00 & 0.00 \\
mean &	0.09 &	12.67&	6.29&	0.00	&12.68&	6.30&	0.08&	0.01&	0.00 &	1.89 \\
3rd Qu. &	0.00&	21.00&	9.00&	0.00&	21.00	&9.00&	0.00	&0.00&	0.00 &	4.29 \\
Max. &	3.00&	29.00&	16.00&	4.00	&30.00&	17.00&	3.00&	6.00&	4.00	&78.57\\
	\end{tabular}
\label{summary}
\end{table}

\subsection{Heatmaps}

\begin{figure}[!htb]
	\centering
	\includegraphics[width=0.31\textheight]{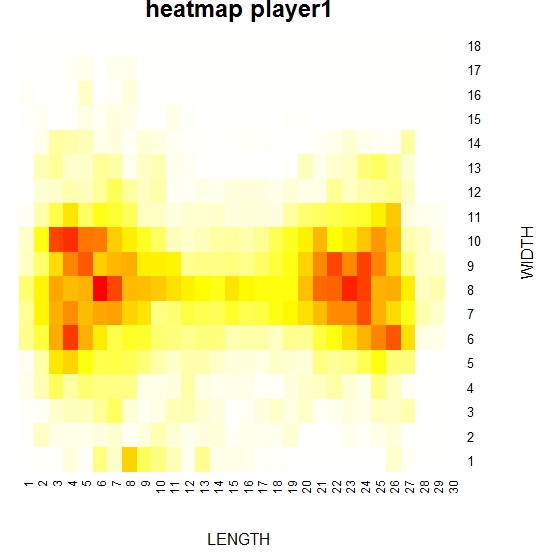}
	\includegraphics[width=0.31\textheight]{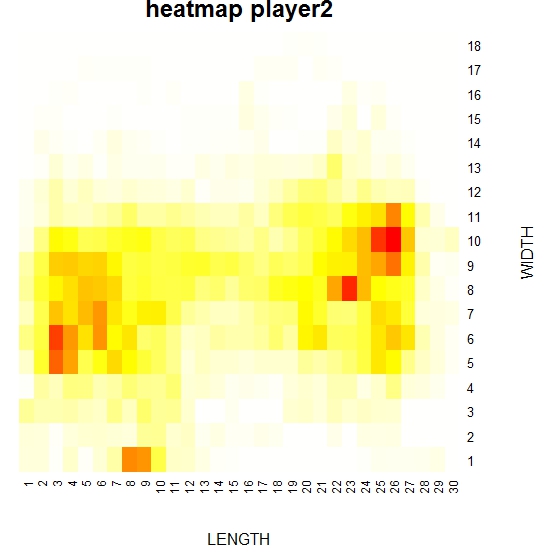}
	\includegraphics[width=0.31\textheight]{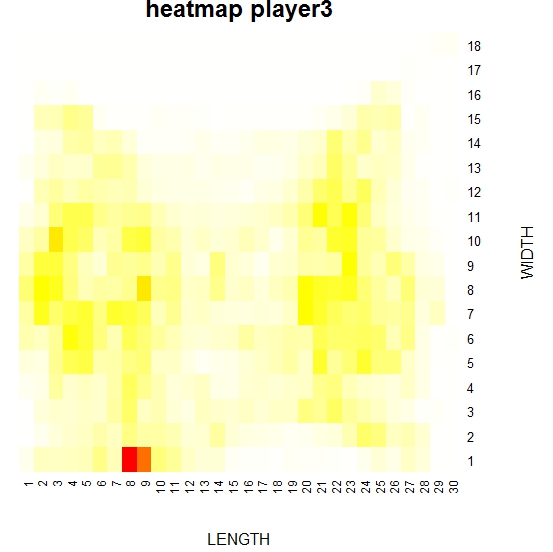}
	\includegraphics[width=0.31\textheight]{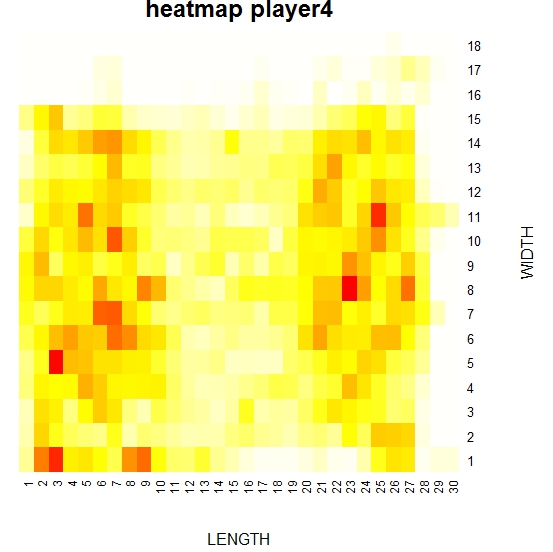}
	\includegraphics[width=0.31\textheight]{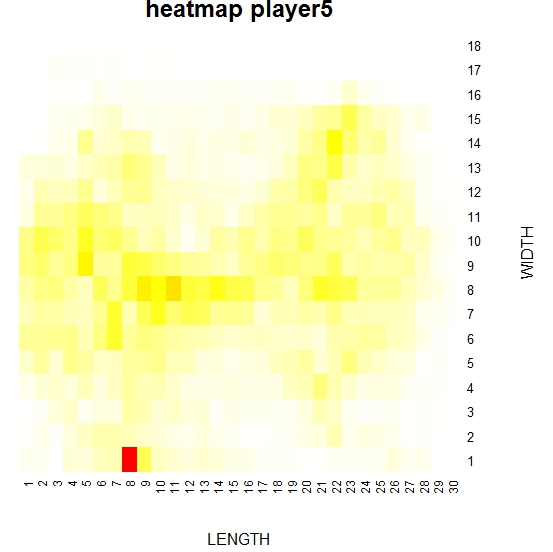}
	\includegraphics[width=0.31\textheight]{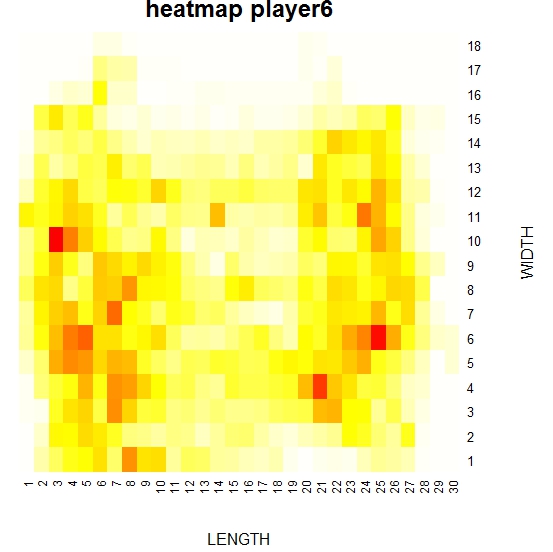}
	\caption{Heatmaps for the six players, in comparison.}
	\label{heat}
\end{figure}

I split the dataset in six smaller datasets each one referring to the location of one of the six players (please refer to appendix \ref{AppB}). I subdivide the playing area into squares of equal size (1 $m^2$) and I count the number of time a player lies in that square.  To do this, I create six non-squared matrices of dimension 15 x 28, as in appendix \ref{AppB}. Each cell of each matrix contains the count of times that a certain player was in the related square. Then, using these matrices, I draw the heatmaps using \texttt{heatmap} function within \texttt{stats} package in R. Heatmaps are reported in figure \ref{heat}. In the figures, the length of the court is reported as the x-axis and the court's width is reported as the y-axis. Colors range from white (lowest intensity, i.e. the player rarely locates in that cell) to red (higest intensity, i.e. the player often locates in that cell) while intermediate intensities are marked with a yellow color. 

By the comparison of the heatmaps is possible to see some differences in the preferred location of each player (figure \ref{heat}). Heatmaps of player 1 and player 2 are similar, in the sense that both players tend to prefer areas close to the basket\footnote{The basket is positioned at the coordinate (1,8).}. Players 4 and 6 show a different locational pattern: their heatmap are less concentrated close to the basket and present an higher level of heterogeneity (i.e. we can also find red cells far away from the basket). Heatmaps of player 3 and player 5 present a red cell close to the bench: it means these two players passed lot of time on the bench.  

\begin{figure}[!htb]
	\centering
	\includegraphics[width=0.34\textheight]{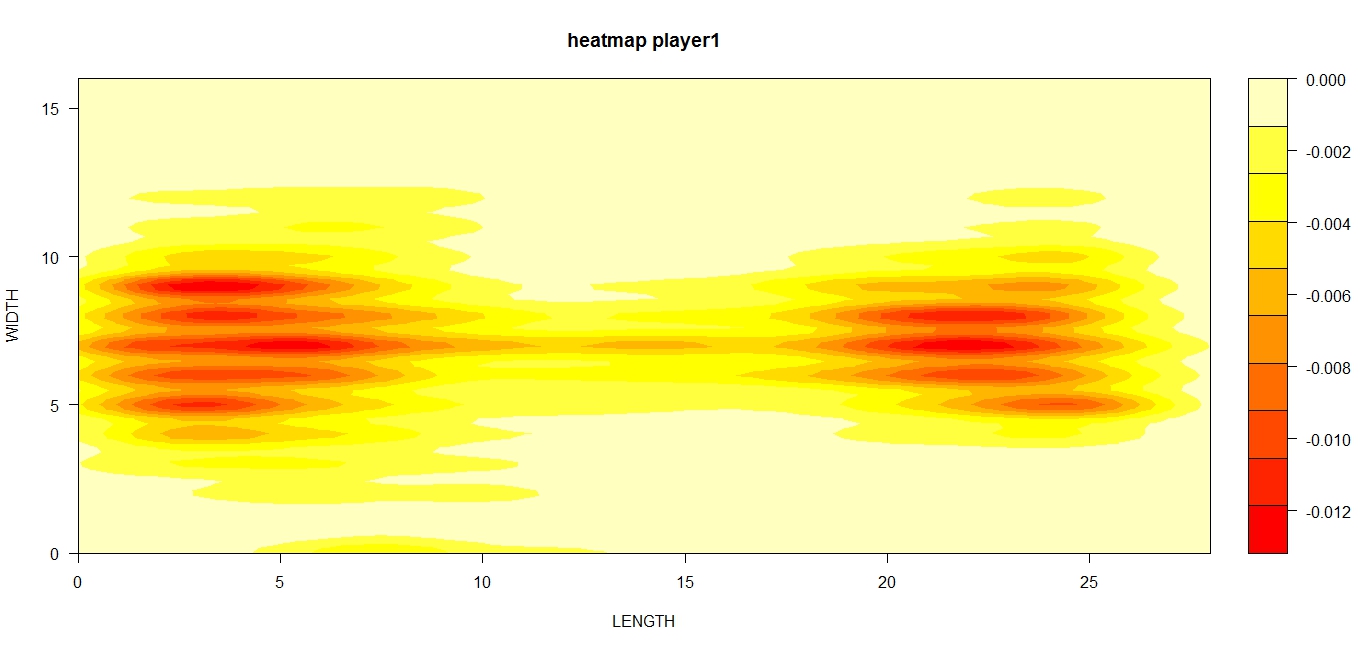}
	\includegraphics[width=0.34\textheight]{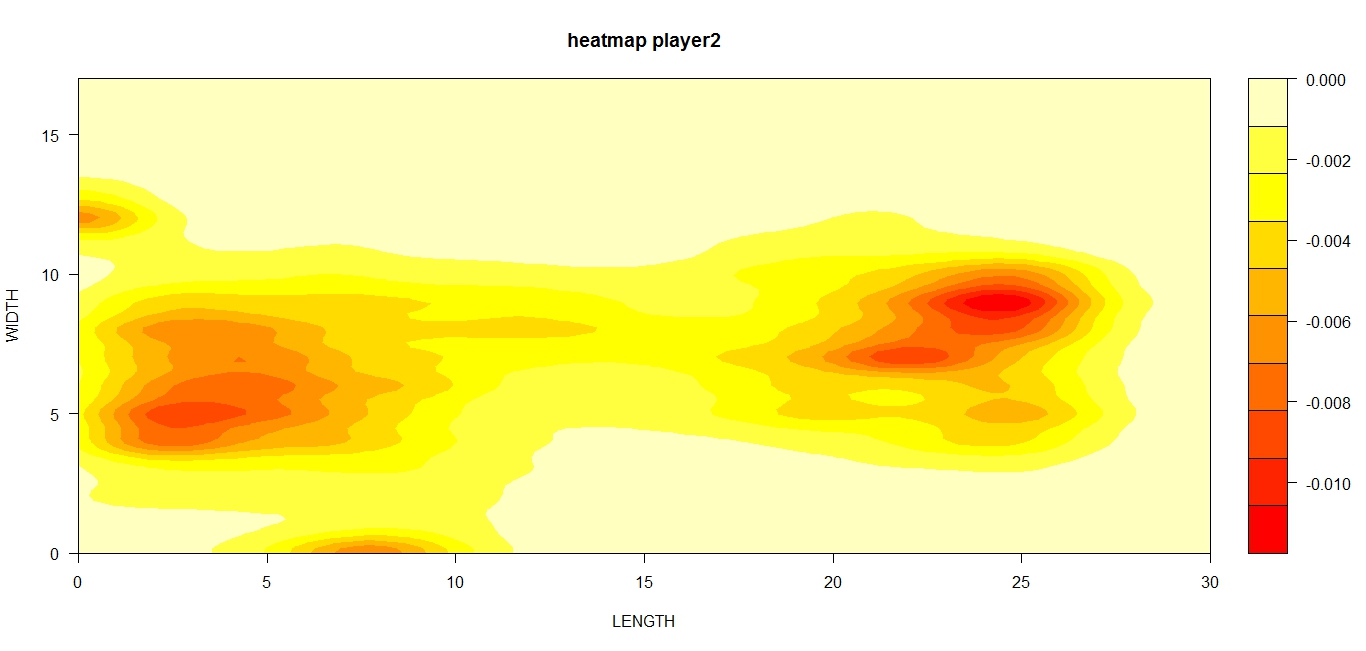}
	\includegraphics[width=0.34\textheight]{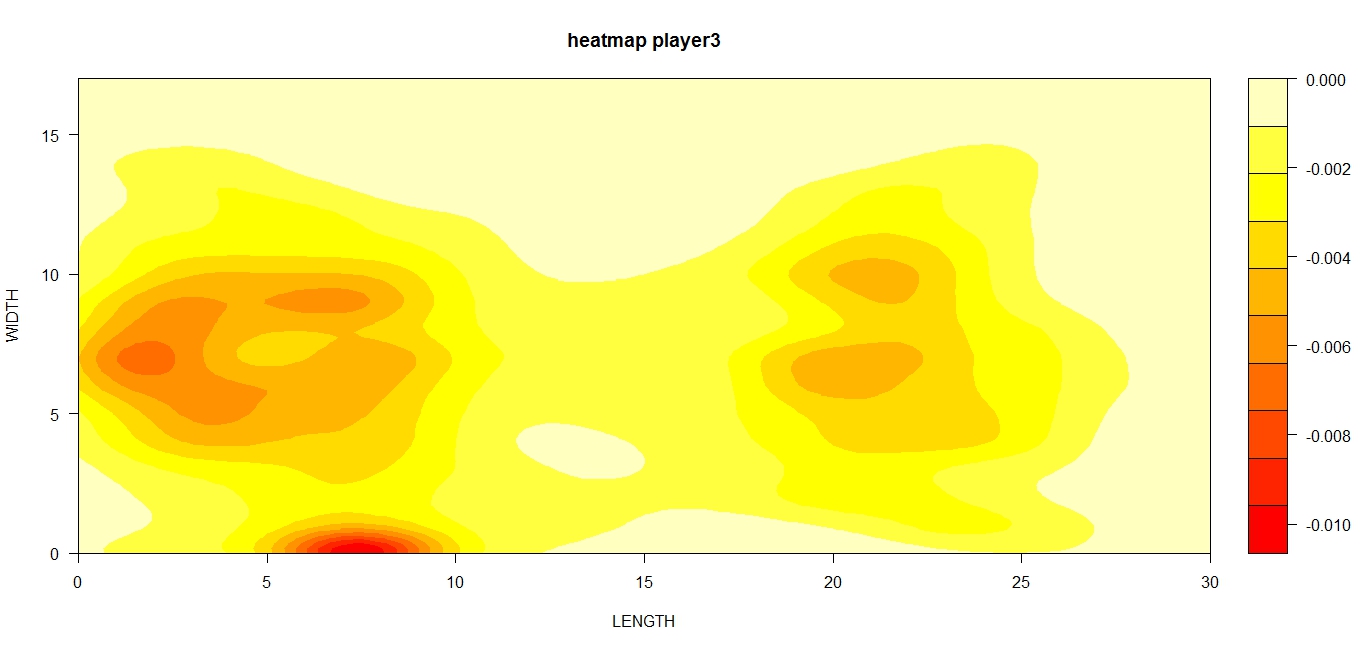}
	\includegraphics[width=0.34\textheight]{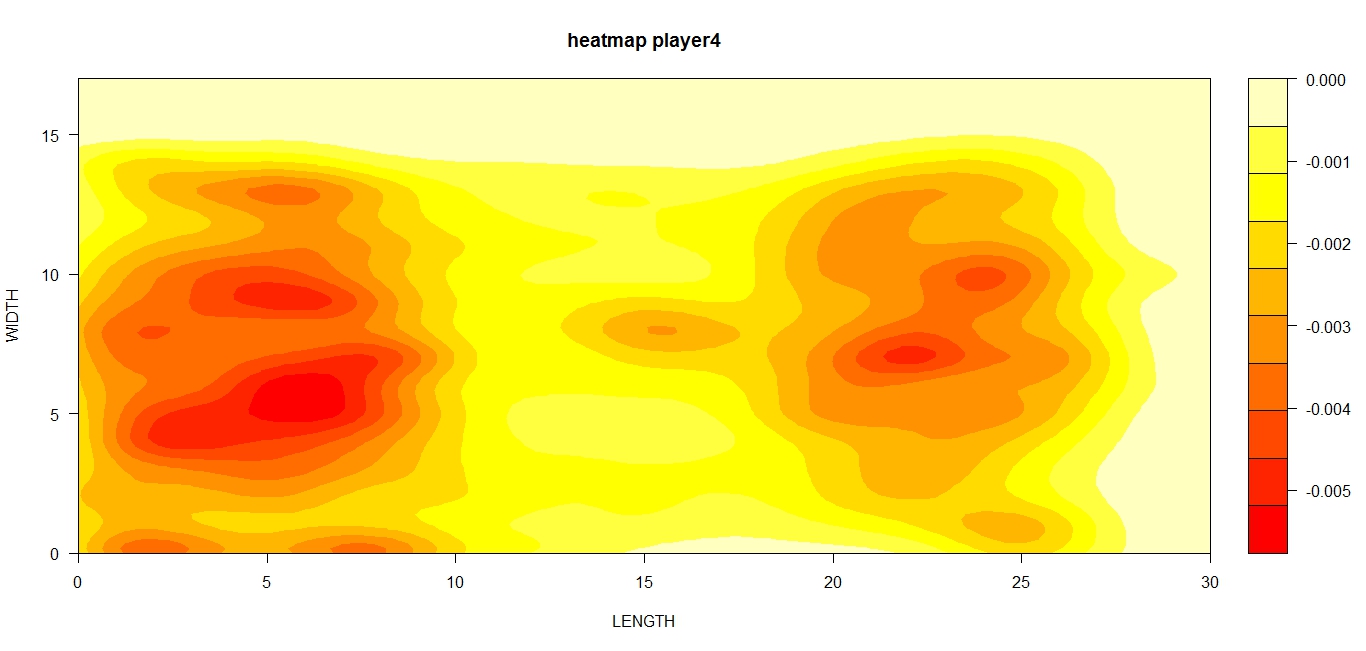}
	\includegraphics[width=0.34\textheight]{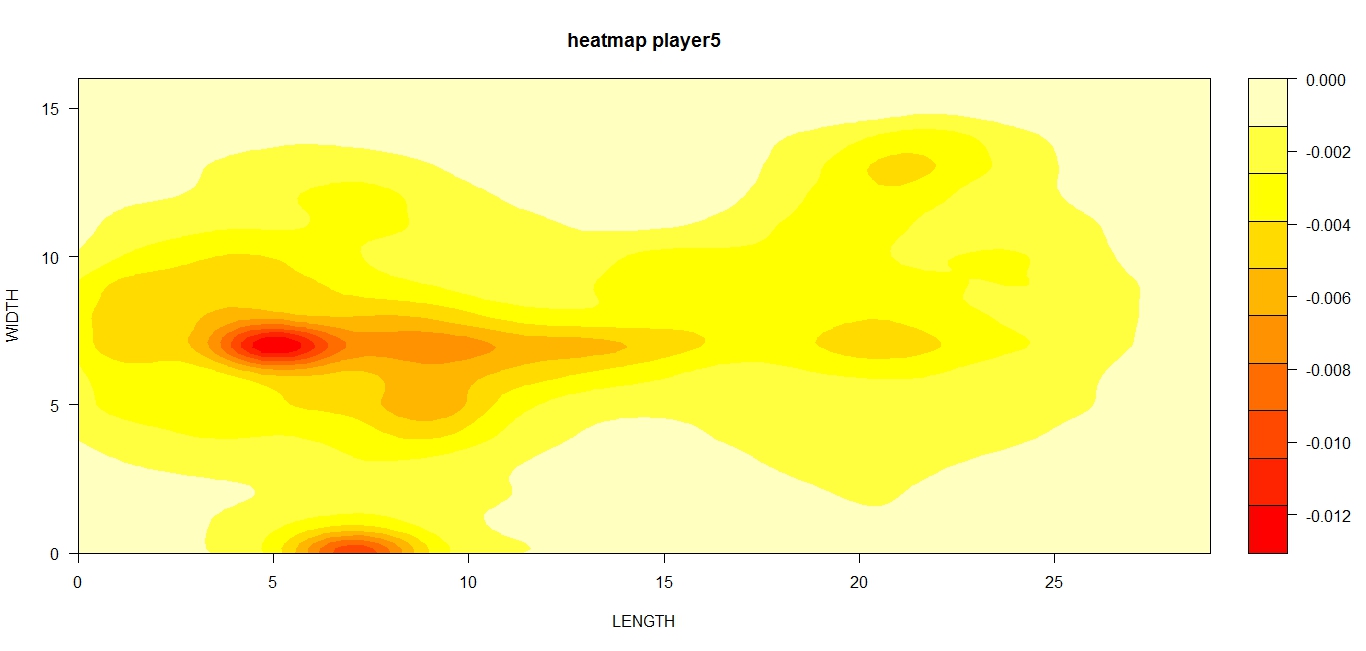}
	\includegraphics[width=0.34\textheight]{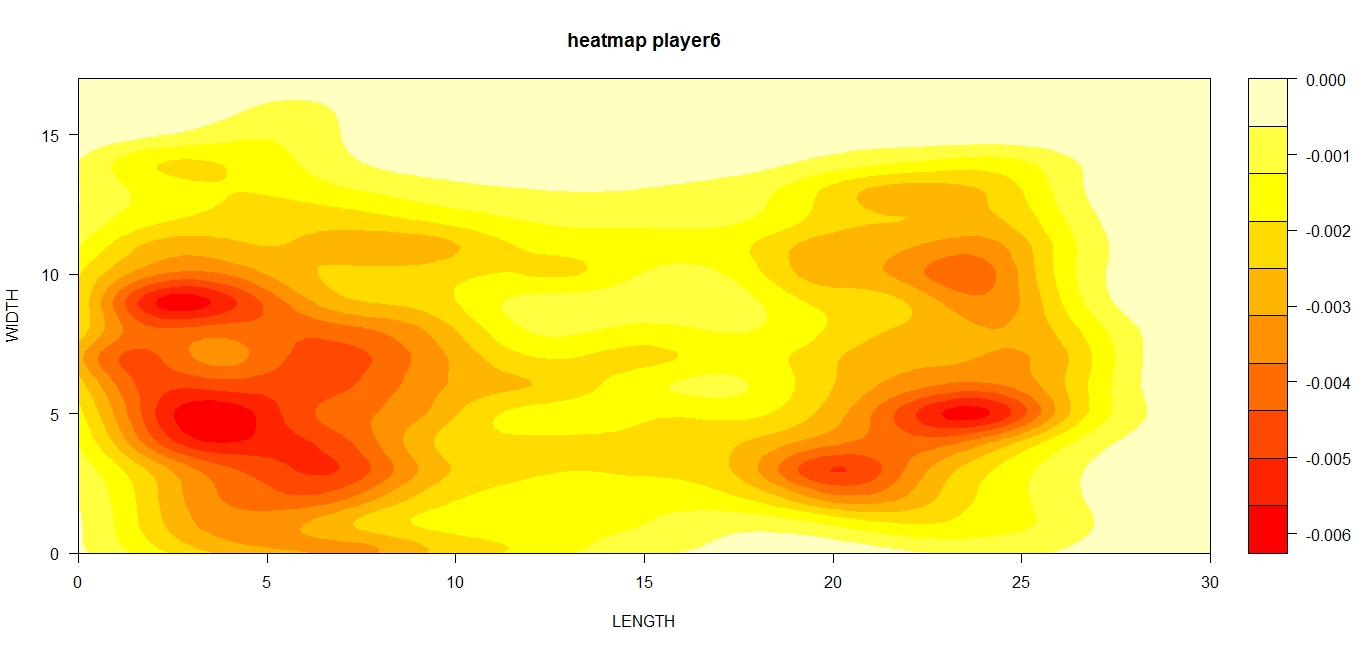}
	\caption{Heatmaps for the six players, in comparison.}
	\label{heat_kernell}
\end{figure}

A Kernel approach is also used here (please refer to appendix \ref{AppC} for codes). A Kernel density estimation (KDE) is a non-parametric way to estimate the probability density function of a random variable. In other words, I replace the exact count of times the players lie in a cell with an estimation of it. I change the rectangular area from a collection of marked cells to a continuous space in which every single point in the area have a certain estimated value. 
Heatmaps generated from the KDE values are reported in figure \ref{heat_kernell}. As for the charts in figure \ref{heat}, the length of the court is reported as the x-axis and the court's width is reported as the y-axis, while colors range from white (lowest density) to red (higest density). These figures provide similar results to previous ones.

\subsection{Motion Charts}

Despite the heatmaps give some hints about the pattern of positioning of players, no conclusions can be done about their dynamic over time and about the interaction among them. The heatmaps completely disregard the time dimension: it is not possible to say something about the location of a certain player in a specific moment or to examine their trajectories. 
Moreover, heatmaps do not shed light on the interactions among players.   

\begin{figure}[!htb]
	\centering
	\includegraphics[width=0.65\textheight]{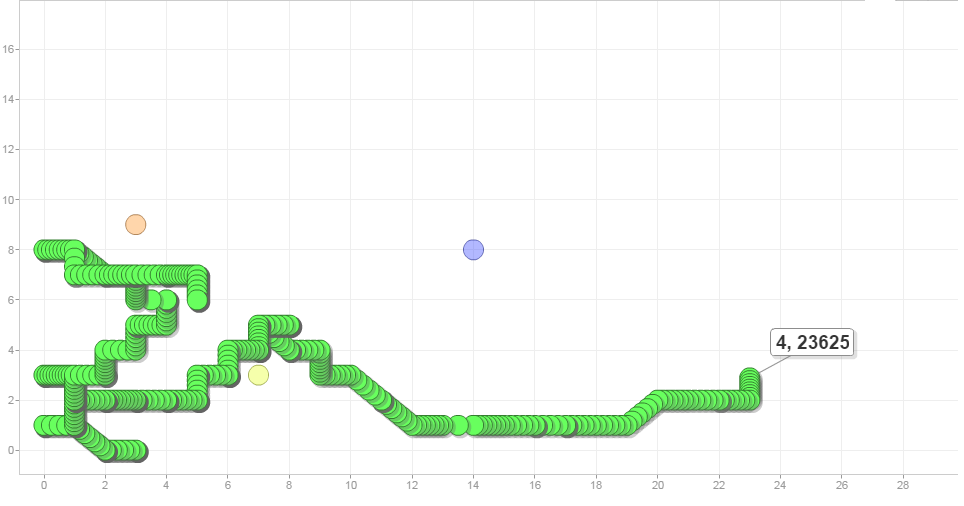}
	\includegraphics[width=0.65\textheight]{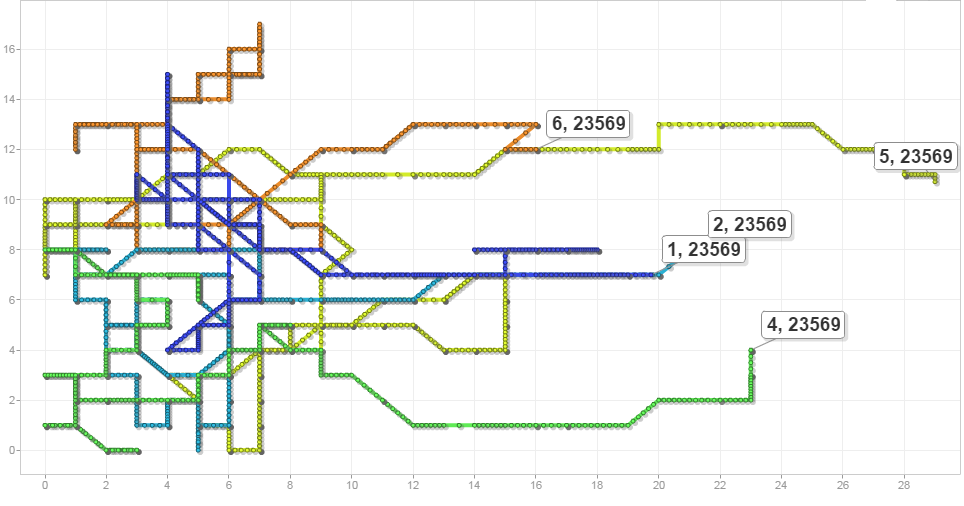}
	\caption{Motion Chart. Example of trail for the movements of player4 (top) and all the five players together (bottom) during an offensive action.}
	\label{trail}
\end{figure}

With motion charts we can account for the time dimension and trace the trajectories of players. This tool allows to analyze the movement of a single player over time as well as the interaction of all the players together (please, refer to Appendix \ref{AppD} to find codes to reproduce the chart). A video showing how motion charts works in our dataset can be found at the link: \url{http://bodai.unibs.it/BDSports/Ricerca2%20-%20DataInn.htm}

Top chart of figure \ref{trail} reports an example of motion chart trail of player 4 during an offensive play.  In this example, the player starts from the bottom right (defensive) region of the court and he moves straight to the left part (the offensive region). Subsequently, he moves first to the bottom and then close to its basket. Player 4 ends its play moving a couple of meters far away from the basket.
A similar analysis could be done by plotting in the same chart the trails of all the five players together (bottom chart of figure \ref{trail}) in order to highlight the interaction among them.

\begin{figure}[!htb]
	\centering
	\includegraphics[width=0.65\textheight]{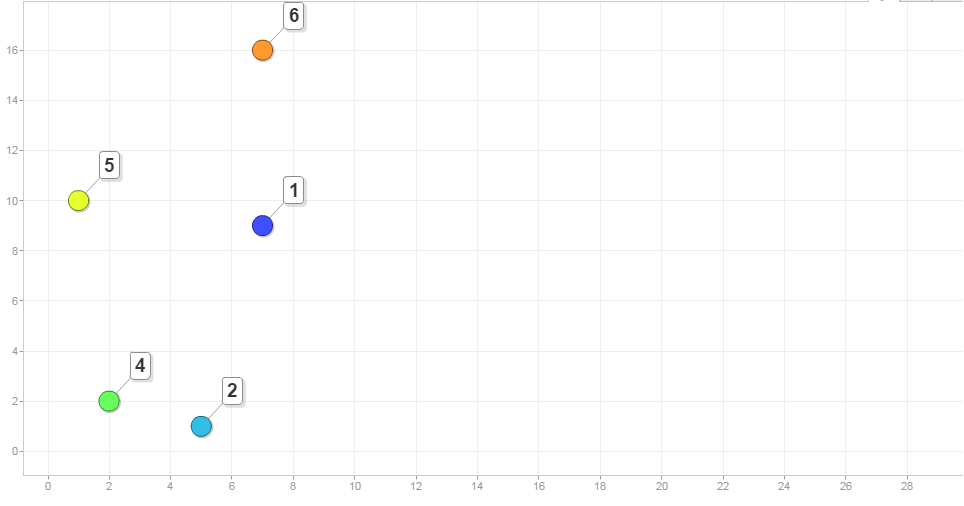}
	\includegraphics[width=0.65\textheight]{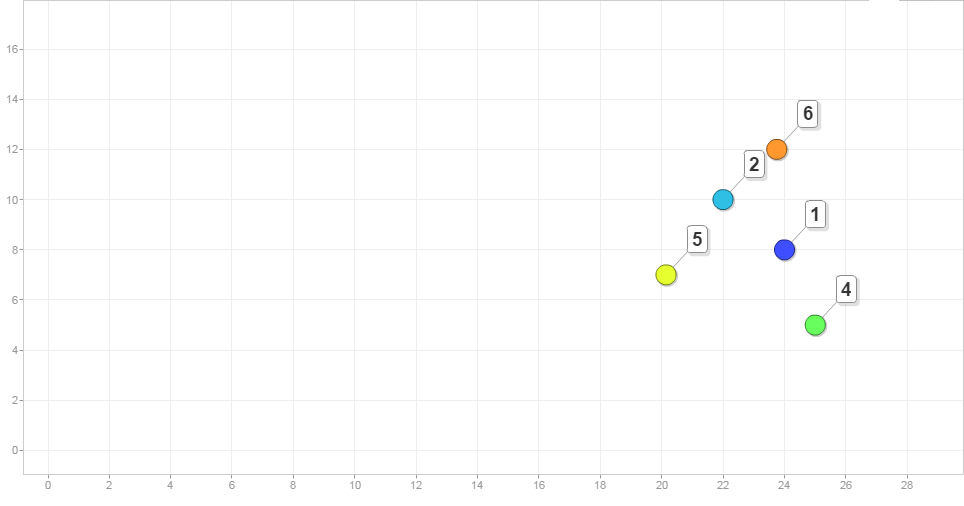}
	\caption{A typical spacing structure for an offensive play (top) and a Typical spacing structure for a defensive play (bottom)}
	\label{MV_A/D}
\end{figure}

Motion charts also allow to analyze the interactions in terms of spacing (i.e. the relative position of a players in terms of the position of the others). A correct spacing could affect the performance of the team. Moreover, to different schemes could correspond different spacing structures. Figure \ref{MV_A/D} reports a typical spacing structure during an offensive play (top) and a typical spacing structure during an defensive play (bottom). It immediately emerges that there is much distance among players during an offensive action. This is not the case of a typical defensive play: in this case, players are really close by.  This makes sense. In fact, in attack, players must be well spaced to effectively play their schemes, move the ball and let the player free to shoot. Conversely, in defense, players may want to play such that the movement of the ball of the opposing team is prevented. In order to do so, players should be positioned close to each other.

\subsection{Further evidences}

The fact that players are more close each other in defensive actions motivate us to further analyze the interaction of players by mean of distances and Voronoi areas. 
 
I compute the average distance for every moment of the match by computing the distance among every pair of players and then by computing the mean of such a distances. Larger is the distance, more spread around the court are the five players. The Voronoi area is computed as the sum of the five Voronoi areas related to each of the five players in the court. Bigger is this value, bigger is the area that players dominate (i.e. that can reach before the players of the opposite team). 
Then I determine, for each moment, whether it was attack or defense, by computing the xy-axes centroid of the location of the five players in the court.  Table \ref{A_D} reports the average distance and the Voronoi area and confirm what the motion chart expressed visually: players are much spread around the court during offensive plays, while they concentrate in closer points when defending. 

\begin{table}[htbp]
	\centering
	\caption{Mean Voronoi area and mean average distance in attack and defense.}
	\small
	\begin{tabular}{l|rr}
		&	Voronoi area ($m^2$) &	Avg. distance (m) \\
		\hline
		Defense		& 28.47 & 5.68 \\
		Attack	& 42.59 & 7.25 	\\
	\end{tabular}
	\label{A_D}
\end{table}

To answer the question if different quintets plays in a different way, and, in doing so, they are more (less) spread in the court, I compute the Voronoi area and the average distance for each possible quintet. There are six quintets when six players rotate: the first is composed by all the players except the player 1, the second quintet is composed by all the players except the player 2, and so on and so forth. Table \ref{quint} reports the Voronoi area and the average distance for each quintet, splitting by attack and defense. Once more, results show that,  when attacking, the team presents a bigger Voronoi area and a smaller average distance among players. When the player 1 is on the bench, both the Voronoi area and the average distance are larger in attacking actions. In defense, the quintets having player 4 or player 5 on the bench play more closer each other; this is confirmed both by the Voronoi area and by the average distance.    

\begin{table}[htbp]
	\centering
	\caption{Mean Voronoi area and mean average distance for different quintets.}
	\small
	\begin{tabular}{l|rr|rr}
		& Attack & & Defense & \\
Quintet		&	Voronoi area &	Avg. distance & Voronoi area &	Avg. distance  \\
		\hline
		1		& 47.54 & 7.70 & 30.52 & 5.85 \\
		2	& 41.09 & 7.09 & 28.97 & 5.76  	\\
		3 & 42.76 & 7.29  & 29.14  & 5.57 \\
		4 & 39.69& 7.13 & 23.00 & 5.03 \\
		5 & 41.46 & 7.04 & 22.99 & 4.99 \\
		6 & 43.03 & 7.15 & 31.22 & 6.22 \\
		
	\end{tabular}
	\label{quint}
\end{table}

Another aspect is to analyze the relation between the distance between players and the team performance. In other words, we want to ask the question whether to play more spread around the court is positively correlated with a good shooting percentage. I match our movements \textit{data.frame} with the play by play and I associate to each minute the two point and three point shoots percentage. Doing this match, I can find the average distance and the Voronoi area in different periods of te match. In detail, I split the match in minutes where the shooting percentage was 0\%, 25\%, 33\%, 50\%, 67\&, 100\%. However, results (Table \ref{perc}) do not show significant differences among groups. 

\begin{table}[htbp]
	\centering
	\caption{Mean Voronoi area and mean average distance in attack, for different moments of the match.}
	\small
	\begin{tabular}{l|rr}
\%		&	Voronoi area ($m^2$) &	Avg. distance (m) \\
		\hline
		0		& 40.90 & 7.18 \\
		25	& 48.95 & 7.48 	\\
		33 & 46.05 &  7.38  \\
		50 & 46.98  & 7.35  \\
	66 & 46.90 & 7.71 \\	
	100  & 40.84 & 6.99 \\
	\end{tabular}
	\label{perc}
\end{table}

\section{Conclusions} \label{sec:concl}

There is a variety of methods and models used in the field of spatio-temporal analysis for team sports that borrow from many research communities, including machine learning, network science, GIS, computational geometry, computer vision, complex systems science and statistics. Analyzing the relation between the team performance and the players' trajectories is a tricky task. To choose the ad-hoc methodology is of vital importance and the availability of a visualization method that guides researchers on this choice it's urgent. 

In this paper I show that \texttt{MotionChart} function from \texttt{googleVis} package in R is a useful tool for visualizing trajectories. The chart properly shows, in time motion, the synchronized trajectories of more then one player on the same 2-dimensional chart. I recommend the use of the motion chart that could be useful in supporting researcher on preliminar stages of their analysis and to facilitate the interpretation of their related results.  With a case study based on basketball player's movements, I show how the tool of the motion charts suggest the presence of interaction among players as well as specific patterns of movements. Guided by these evidences, I have computed Voronoi areas and distances among players for offensive and defensive actions, and for different quintets in the court. Evidences suggested by the motion charts have been confirmed. 

Future developments aim to adopt spatial statistics and spatial econometrics techniques applied to trajectory analysis \cite{brillinger2008modelling}, such as bivariate K-function method \cite{arbia2008class}. Adapting these techniques to team sports will contribute in this field by better characterizing specific patterns of players' movements.    

\section*{Acknowledgments}

Research carried out in collaboration with the Big\&Open Data Innovation Laboratory (BODaI-Lab),
University of Brescia (project nr. 03-2016, title \textit{Big Data Analytics in Sports}, \url{www.bodai.unibs.it/BDSports/}), granted by Fondazione Cariplo and Regione Lombardia.
I would like to thanks  Paola Zuccolotto and Marica Manisera (University of Brescia) for giving me suggestions during the preparation  of this paper. Furthermore, I thanks Tullio Facchinetti and Federico Bianchi (University of Pavia) for having helped me with the data interpretation.

\section*{Appendices}
\addcontentsline{toc}{section}{Appendices}
\renewcommand{\thesubsection}{\Alph{subsection}}
\subsection{Codes for preparing the \textit{data.frame}} \label{AppA}

\texttt{
	\#upload the data.frame\\
	>ds = read.delim("dataset.txt") \\
	>str(ds) \\
	\\
    'data.frame':   133662 obs. of  17 variables: \\
	\$ id             : int  1130547 1130548 1130549 1130550 1130551 1130552 1130553 1130554 1130555 1130556 ...\\
	\$ insert\_date    : Factor w/ 101 levels "24/03/2016 19:01",..: 1 1 1 1 1 1 1 1 1 1 ...\\
	\$ tagid          : Factor w/ 6 levels "84eb18675b32",..: 3 3 3 3 3 3 3 3 3 3 ...\\
	\$ position\_ts    : Factor w/ 101 levels "24/03/2016 19:01",..: 1 1 1 1 1 1 1 1 1 1 ...\\
	\$ smt\_x          : int  0 0 0 0 0 0 0 0 0 0 ...\\
	\$ smt\_y          : int  -1 -1 -1 -1 -1 -1 -1 0 0 0 ...\\
	\$ smt\_z          : int  -2 -2 -1 -1 -1 -1 -1 -1 -1 -1 ...\\
	\$ klm\_x          : int  0 0 0 0 0 0 0 0 0 0 ...\\
	\$ klm\_y          : int  -1 -1 0 0 0 0 0 0 0 0 ...\\
	\$ klm\_z          : int  -1 -2 -1 -1 -1 -1 -1 -1 -1 -1 ...\\
	\$ klv\_x          : int  0 0 0 0 0 0 0 0 0 0 ...\\
	\$ klv\_y          : int  0 0 0 0 0 0 0 0 0 0 ...\\
	\$ klv\_z          : int  0 0 0 0 0 0 0 0 0 0 ...\\
	\$ tagid\_new      : int  3 3 3 3 3 3 3 3 3 3 ...\\
	\$ time           : int  2 1 3 4 5 6 7 8 9 10 ...\\
	\$ speed.mtr.sec  : num  5.29 NA 4.81 0 0 ...\\
	\$ timestamp\_ms\_ok: num  42497841 42497652 42498049 42498271 42498374 ...\\
	\\
	\#upload an instrumental data.frame\\
	>id = read.delim("id\_matrix.txt") \\
	>str(id) \\
	\\
	'data.frame':   540 obs. of  2 variables: \\
	\$ long : int  0 0 0 0 0 0 0 0 0 0 ...\\
	\$ short: int  0 1 2 3 4 5 6 7 8 9 ...\\
	\\
	\# install packages\\
	>install.packages("googleVis") \\
	>library(googleVis) \\
	>install.packages("stats") \\
	>library(stats) \\
	>install.packages("MASS") \\
	>library(MASS) \\
	\\
	\#select only positive value (inside the court) \\
    >ds = ds[ds\$klm\_y >=0 \& ds\$klm\_z >= 0,] \\
    \#remove the pre-match period \\
    >ds = ds[ds\$timestamp\_ms\_ok >= 47319496,] \\
    \#remove the half-time break \\
    >ds = ds[ds\$timestamp\_ms\_ok >= 49134256 | ds\$timestamp\_ms\_ok <= 49021530,] \\ 
    \#remove the post-match period\\
>ds = ds[ds\$timestamp\_ms\_ok <= 51377646,] 
}

\subsection{Codes for producing Heatmaps} \label{AppB}

\texttt{
	 \# generate a dataset for each player \\
	player1 = ds[ds\$tagid == "84eb18675b32",] \\player2 = ds[ds\$tagid == "84eb18675b4b",] \\ player3 = ds[ds\$tagid == "b4994c898155",] \\ player4 = ds[ds\$tagid == "b4994c89889c",] \\ player5 = ds[ds\$tagid == "b4994c8baa73",] \\ player6 = ds[ds\$tagid == "b4994c8bcc29",] \\ 
	\\
	 \# create a data.frame for producing the heatmaps\\ 
	 player1\$conc = paste(player1\$klm\_y, player1\$klm\_z, sep = ",")	\\
	player1\$count = 1
	\\player1\_agg = aggregate(player1\$count, by = list(player1\$conc),FUN=sum, \\ na.rm=TRUE)
	\\dim(player1\_agg)
	\\names(player1\_agg) = c("position\_long\_short","freq")
	\\sum = sum(player1\_agg\$freq)
	\\player1\_agg\$rel\_freq = player1\_agg\$freq/sum
	\\player1\_agg[order(-player1\_agg\$freq),][1:50,]
	\\prepare the aggregate dataset to the heat map
	\\str =  strsplit(player1\_agg\$position\_long\_short,",")
	\\mat  = matrix(unlist(str), ncol=2, byrow=TRUE)
	\\df   = as.data.frame(mat)
	\\colnames(df) = c("long", "short")
	\\player1\_agg  = cbind(player1\_agg, df)
	\\player1\_agg\_bal = merge(id,player1\_agg, by.x=c("long","short"), \\ by.y=c("long","short"), all.x=TRUE)
	\\dim(player1\_agg\_bal)
	\\player1\_agg\_bal\$freq[is.na(player1\_agg\_bal\$freq)] = 0
	\\matrix\_player1 = matrix(player1\_agg\_bal\$freq,18,30)
	\\max(matrix\_player1) \\
	\\
	\# producing the heatmap for the player 1 \\
	heatmap\_player1 = heatmap(-matrix\_player1, Rowv=NA, Colv=NA, \\ col = heat.colors(max(matrix\_player1), alpha = 1), scale="none",\\ xlab = "LENGTH", ylab =  "WIDTH", main = "heatmap player1")
}

\subsection{Codes for the Heatmap with Contour plot} \label{AppC}

\texttt{
	\\m1 = player1[,9:10]
	\\dens = kde2d(m1\$klm\_y, m1\$klm\_z, n=100)
	\\dens\$z = -dens\$z 
	\\i = min(dens\$z)
	\\min = 0.00001
	\\int = c(i, i*0.9, i*0.8, i*0.7, i*0.6, i*0.5, i*0.4, i*0.3, i*0.2, i*0.1,min)
	\\
	\\filled.contour(dens, col = heat.colors(10, alpha = 1), levels = int, xlab = "LENGTH", ylab =  "WIDTH", main = "heatmap player2")}

\subsection{Codes for the googleVis Motion Chart} \label{AppD}

\texttt{MC = gvisMotionChart(ds, idvar = "tagid\_new", timevar = "time", xvar = "klm\_y", yvar = "klm\_z", colorvar = "", sizevar = "", options=list(width=1200, height=600))
\\plot(MC)}

\end{document}